\documentclass[useAMS,usenatbib]{mn2e} 
\usepackage{graphicx} 
\usepackage[usenames]{color}

\title[]{Testing different methods for atmospheric parameters determination. The case study of the Am star HD\,71297\thanks{Based 
          on observations made with the Italian Telescopio
          Nazionale Galileo (TNG) operated on the island of La Palma by the Fundación Galileo Galilei of the 
          INAF (Istituto Nazionale di Astrofisica) at the Spanish Observatorio del Roque de los Muchachos of
          the Instituto de Astrofisica de Canarias}}

\author[G. Catanzaro et al.]{G. Catanzaro$^{1}$\thanks{E-mail: gca@oact.inaf.it}, 
V. Ripepi$^{2}$, 
H. Bruntt$^{3}$
\\  
$^{1}$INAF-Osservatorio Astrofisico di Catania, Via S.Sofia 78, I-95123, Catania, Italy\\ 
$^{2}$INAF-Osservatorio Astronomico di Capodimonte, Via Moiariello 16, I-80131, Napoli, Italy\\ 
$^{3}$Department of Physics and Astronomy, Building 1520, Aarhus University, 8000 Aarhus C, Denmark\\
} 
 
\date{Accepted  2013 March 4. Received 2013 March 4; in original form 2012 December 14} 
 
\pagerange{\pageref{firstpage}--\pageref{lastpage}} \pubyear{2002} 
 
\def\LaTeX{L\kern-.36em\raise.3ex\hbox{a}\kern-.15em 
 T\kern-.1667em\lower.7ex\hbox{E}\kern-.125emX} 
 
\begin{document} 
 
\label{firstpage} 
 
\maketitle 
 
\begin{abstract} 
In this paper we present a detailed spectroscopic analysis of the suspected marginal Am star HD\,71297.
Our goal is to test the accuracy of two different approaches to determine the
atmospheric parameters effective temperature, gravity, projected rotational velocity, and chemical
abundances. The methods used in this paper are: classical spectral
synthesis and the {\it Versatile Wavelength Analysis} ({\sl VWA}) software.

Since our star is bright and very close to the Sun, we were able to determine its effective temperature and gravity 
directly through photometric, interferometric, and parallax measurements. The values found were taken as reference
to which we compare the values ​​derived by spectroscopic methods. Our analysis leads us to conclude that the spectroscopic
methods considered in this study to derive fundamental parameters give consistent results, if we consider all the 
sources of experimental errors, that have been discussed in the text. In addition, our study shows that the spectroscopic 
results are quite as accurate as those derived from direct measurements. 

As for the specific object analyzed here, according to our analysis, HD\,71297 has chemical abundances 
not compatible with the previous spectral classification. We found moderate underabundances of carbon, sodium,
magnesium, and iron-peak elements, while oxygen, aluminum, silicon, sulfur, and heavy elements (Z $\geq$ 39) 
are solar in content. This chemical pattern has been confirmed by the calculations performed with both methods.

\end{abstract} 
 

\section{Introduction} 
The current era is characterized by an enormous growth of stellar data
of different nature. The advent of space missions, such as
CoRoT~\citep[Co\emph{nvection}, Ro\emph{tation and planetary}
Tr\emph{ansits};][]{baglin07} and {\it Kepler} \citep{borucki97},
designed to obtain precise photometry for an impressive number of
stars, have led astronomers to undertake several ground-based
spectroscopic campaigns in order to get as accurate as possible
fundamental parameters such as effective temperature, surface gravity,
projected rotational velocity, and metallicity. The analysis of this huge amount of data is
possible, on reasonably short timescales, only through the use of
automatic or semi-automatic procedures.

Among the various targets observed by {\it Kepler}, Am stars are
assuming an ever growing importance.  In fact, it was once thought
that Am stars did not pulsate, and that the explanation for this is
that atomic diffusion is expected to drain helium from the He{\sc ii}
driving zone. More intensive observations have revealed that this
picture is not correct and several Am stars are known to pulsate from
ground-based observations \citep[based on superWASP photometry,
see][]{smalley2011} as well as from {\it Kepler} satellite data
\citep[see][]{balona11}. In this last paper, the authors showed that an
accurate measure of the location of the pulsating Am stars in the HR
diagram is crucial to pin down effectively the pulsation models and to
put observational constraint on the instability strip for pulsation in
Am stars. However, for most of the investigated objects, a modern
determination of the stellar parameters such as effective
temperatures, gravities and chemical abundances, based on high
resolution spectroscopy is still lacking. To fill this gap, we have already
started a spectroscopic campaign devoted to obtain high resolution
spectra of Am stars, being the first data obtained with the spectrograph
SARG installed at the Italian telescope {\it TNG}.  Our purpose in the
near future is to analyze this growing amount of spectra to derive
accurate stellar parameters for the investigated objects in the
shortest possible time. 

There are several examples in the literature of papers regarding various 
methods for automatic (or semi-automatic) spectroscopic analysis, \citep{tkachenko12,
lehmann11,niemczura09,erspamer03}. However, it is important to
ascertain the accuracy of the derived parameters obtained using different methods. A
few recent papers have partially addressed this problem, reporting that
significant differences can be found both in the derived stellar
parameters \citep[see e.g.][]{fossati2011} and/or in the estimated
uncertainties in these quantities \citep[see e.g.][]{joanna2010}.

With the aim of understanding how accurately we can derive the 
stellar parameters and the HR location for our sample of Am stars, we 
decided to concentrate our efforts on comparing the results that can
be obtained by analyzing the high resolution spectrum for one star in
our sample with the two methods available
to us: i) classical spectral synthesis \citep[see][and references
therein]{catanzaro11,catanzaro12} and ii) Versatile Wavelength
Analysis ({\sl VWA}) \citep[see][and references
therein]{bruntt10a,bruntt10b}. 

To carry out this exercise, we choose to analyze in detail the Am star
HD\,71297. This is the most suitable object in our sample because it is 
very bright (V\,=\,5.58 mag) and the SARG@TNG spectrum has a very good
signal-to-noise ratio. Moreover, HD\,71297's brightness allowed us to
find very useful observational data in the literature, including an accurate
{\it Hipparcos} parallax and interferometric measures which allow us
to estimate independently from spectroscopy the stellar parameters.   
Thus, the main goal of this paper is to ascertain if the two quoted
 methods of quantitative spectral analysis lead to different or, on the contrary, 
to comparable results when used to exploit exactly the same
observational material.

\section{The target. The Am: star HD\,71297} 

HD\,71297 has been classified as suspected marginal Am star by
\citet{cowley68}, with a spectral type of A5m: (colon denoted its {\it
marginal} nature). We remind readers the definition of marginal Am star 
as an Am star in which there is a difference of less than five subclasses 
between the K-line and the metal line spectral types, and in which the line strength
anomalies are mild. \citet{cowley68} concluded that the star exhibits
only weak metallic lines, looking like the spectrum of
$\beta$~Ari. This classification has been confirmed one year later by
\citet{cowley69}.

\citet{abt85} searched for binarity among a sample of 60 Am stars and
concluded that HD\,71297 may be variable in radial velocity with
a period of a hundred days. They assigned the spectral type of kA8hA9mF0
to HD\,71297.

This star has also been studied by \citet{guthrie87} who derived
atmospheric parameters, T$_{\rm eff}$\,=7900~K, $\log g$=4.2, and
calcium abundance, $\log N({\rm Ca})$\,=\,6.33, expressed in a scale
in which $\log N(H)$\,=\,12.0, that is practically the solar
value. Later on, \citet{kuenzli98} revisited the star, finding it a
little bit cooler and less evolved, T$_{\rm eff}$\,=7712~K and $\log
g$=4.06, but still with solar calcium abundance.

No other studies have been found in the recent literature, especially
no detailed analysis of the chemical pattern of its atmosphere.  Thus, we
chose this object as a benchmark for the two methods of analysis we
would like to compare in this paper.
  
\begin{table*}
\caption{Evaluations of T$_{\rm  eff}$ and $\log g$ values of
  HD\,71297 on the basis of various photometric systems (see text for
  details). }
\label{tabPhot} 
\centering         
\begin{tabular}{lccc}
\hline
\hline
Photometry   &  T$_{\rm  eff}$ &  $\log g$ & Calibration       \\
\hline
$uvby$--$\beta$ & 7700$\pm$40 & 4.06$\pm$0.04 & \citet{napi93} \\
$uvby$--$\beta$ & 7970$\pm$30 & 4.28$\pm$0.04 & \citet{ribas97} \\
$UBVB_1B_2V_1G$ & 7770$\pm$80 & 4.46$\pm$0.06 & \citet{kuenzli98} \\
$uvby$--$\beta$ & 7770$\pm$125 & 4.08$\pm$0.125 & \citet{sk97} \\
$uvby$--$\beta$ & 7800$\pm$125 & 4.06$\pm$0.125 & \citet{heiter02} MLT $\alpha$=0.5 \\
$uvby$--$\beta$ & 7780$\pm$125 & 4.06$\pm$0.125 & \citet{heiter02} using \citet{cgm96}
\\
\hline
\hline
\end{tabular}
\end{table*}

\section{Observation and data reduction}
\label{obs}

Spectroscopic observations of HD\,71297 were carried out with the SARG
spectrograph, which is installed at the {\it Telescopio Nazionale
  Galileo}, located in La Palma (Canarias Islands, Spain).  SARG is a
high-resolution cross-dispersed echelle spectrograph \citep{gratton01}
that operates in both single-object and longslit (up to 26) observing
modes and covers a spectral wavelength range from 370~nm up to about
1000~nm, with a resolution ranging from R = 29\,000 to 164\,000.

Our spectra were obtained on 2011, February 21 at R\,=\,57\,000 using
two grisms (blue and yellow) and two filters (blue and yellow).  These
were used in order to obtain a continuous spectrum from 3600 {\AA} to
7900 {\AA} with significant overlap in the wavelength range between
4620 {\AA} and 5140 {\AA}. We acquired the spectra for the star with
an exposure time of 120~sec. and a signal-to-noise ratio S/N of at
least 100 per pixel in the continuum.  For example, in the region
centered around the Mg{\sc i} triplet at 5170\,{\AA} the measured S/N
is about 120.

The reduction of all spectra, which included the subtraction of the
bias frame, trimming, correcting for the flat-field and the scattered
light, the extraction for the orders, and the wavelength calibration,
was done using the NOAO/IRAF packages\footnote{IRAF is distributed by
  the National Optical Astronomy Observatory, which is operated by the
  Association of Universities for Research in Astronomy, Inc.}. The
IRAF package {\it rvcorrect} was used to make the velocity corrections
due to Earth's motion to transform the spectra to the barycentric rest
frame.

\section{Stellar parameters from photometry, interferometry and parallax.}
\label{parallax}

Before  starting with the analysis of the spectrum of HD\,71297, it is
extremely important to estimate the stellar parameters, mainly T$_{\rm
  eff}$ and $\log g$, through independent methods, such as those based
on intermediate-band photometry, parallax, interferometry etc. This
is useful to have: i) a reliable starting point for the spectroscopic
analysis and ii) a trustworthy and accurate independent evaluation of
the stellar fundamental parameters to compare with.

\subsection{Evaluation of T$_{\rm eff}$ and $\log g$ from Str\"omgren
  and Geneva photometry}
\label{phot_sect}

A first estimate of T$_{\rm eff}$ and $\log g$ for HD\,71297 can be
obtained from the Str\"omgren photometry: $V=5.607\pm0.002$,  $b-y = 0.123\pm0.002$, 
$m_1 = 0.197\pm0.004$, $c_1 = 0.833\pm0.006$, $\beta = 2.831\pm0.003$
\citep{hauck98}. However, It is important to check the value of $V$-band magnitude because
HD\,71297 is a (low-amplitude) variable star \citep[see][]{balona11}. A safe average
$V$ magnitude can be derived from the time-series photometry in the
Hipparcos ($H_p$) and Tycho  ($V_T$) photometric systems. Including the
uncertainties in the trasformations to the Johnson $V$ system
\citep{bessell2000}, we get $V$=5.59$\pm$0.01 mag and $V$=5.60$\pm$0.01
mag. Finally, an additional independent measure was reported by \citet{kuenzli98}
who measured $V$=5.604$\pm$0.008 mag. On this basis, we adopted a
value of $V$=5.60$\pm$0.01 mag for HD\,71297. This value is consistent
within the errors with all the above quoted evaluations.
 
The other Str\"omgren indices should be less uncertain with respect to $V$-band, however it is likely that the
errors are somewhat underestimated. 

We adopted an updated version of the {\it TempLogG}\footnote{available
  through http://www.univie.ac.at/asap/manuals/\\tipstricks/templogg.localaccess.html}
software \citep{rogers95} to estimate T$_{\rm eff}$ and $\log g$ by using the calibrations present  in
the package. We did not consider the older calibration by \citet{balona84} and \citet{moon85,md85}, 
but only the more recent ones by \citet{napi93,ribas97,kuenzli98}. 
Note that the last calibration is for the Geneva $UBVB_1B_2V_1G$ system, whose
photometry for  HD\,71297 is provided by \citet{rufener88}.
These results are summarized in the first three rows of
Table~\ref{tabPhot}. In addition, we considered the results by \citet{sk97} and \citet{heiter02} 
who provided $uvby$ grids based on the Kurucz model atmospheres but  with different
treatment of the convection. In particular we report on
Table~\ref{tabPhot} the T$_{\rm eff}$ and $\log g$ estimated by interpolating the \citet{sk97} grids
that were built using the \citet{cm91} convection
treatment. Similarly, the last two column of Table~\ref{tabPhot}  list
the stellar parameters obtained using two choices for the grids\footnote{These grids
  are available on the NEMO site  www.univie.ac.at/nemo/gci-bin/dive.cgi} by \citet{heiter02}: 
i) standard mixing-lenght theory (MLT)\footnote{defined as the ratio  {\it $\alpha$=l/H$_p$} of 
convective scale length {\it l} and local pressure scale height {\it H$_p$}} with $\alpha$=0.5; ii)
the \citet{cgm96} treatment of the convection. In all these cases the
error associated to the parameters was imposed as half-the width of
the grid.
An inspection of Table~\ref{tabPhot} reveals
some dispersion among the various calibrations both in 
T$_{\rm  eff}$ and $\log g$, even if the last three values are
absolutely equivalent within the errors. Quantitatively, a simple average of the results gives:
$\langle$T$_{\rm  eff}$$\rangle$=7800$\pm$90 K and $\langle$$\log g$$\rangle$=4.17$\pm$0.17 dex.

As for the interstellar reddening, {\it TempLogG} gives as a result
$E(b-y)$=0.009$\pm$0.003 mag. This value is probably
non-significant, because of the very low value of both reddening and
associated uncertainty (underestimated in our opinion). 
Since the star is only 50 pc far from the Sun
(see next sub-section), it is acceptable to neglect the effect of the
interstellar absorption. This assumption was also confirmed a
posteriori by looking into the spectrum of HD\,71297 for the 
interstellar Na{\sc i} lines at 5890.0, 5895.9 {\AA} which can be 
used to estimate the interstellar reddening
\citep[see][]{munari97}. Indeed, these lines are practically
undetectable. To conclude, in the following analysis we have neglected
the interstellar reddening.

\subsection{Evaluation of the stellar parameters from Photometry,
  Parallax and Interferometry}

The {\it Hipparcos} parallax for HD\,71297, $\pi$\,=\,19.99\,$\pm$\,0.38\,mas
\citep{leeuwen}, allows us to estimate with the excellent accuracy of 2\% the
distance of this object: d\,=\,50\,$\pm$\,1.0~pc. It is then
straightforward to calculate the visual absolute magnitude of 
the target: $M_V = 2.105 \pm 0.045$ mag. To derive the luminosity we
need to evaluate first the visual bolometric correction BC$_V$. To this aim
we adopted the models by \citet{bessel98} where it is assumed that 
M$_{\rm bol,\odot}$\,=\,4.74 mag\footnote{To take into account the
uncertainty on the bolometric magnitude of the Sun, in all the
calculations we associated an error of 0.01 mag to this value}. 
We then get BC$_V$\,=\,0.03\,$\pm$\,0.01 mag for HD\,71297. We note
that to obtain this value we interpolated the \citet{bessel98} tables 
by using as input the T$_{\rm  eff}$ and $\log g$ obtained from the photometry. The error was
conservatively assumed to be 0.01 mag to take into account the
observed dispersion of the photometrically derived T$_{\rm  eff}$ and $\log
g$. We have now estimated all the quantities needed to calculate the
luminosity of  HD\,71297: $\log(L/L_\odot) = 1.04 \pm
0.02$\footnote{Whenever possible, we have summed in quadrature the
  uncertainties on the single quantities when propagating the errors}.

\begin{table}
\caption{Astrophysical quantities for HD\,71297 evaluated
  independently from spectroscopy.}
\label{phot} 
\centering         
\begin{tabular}{lr}
\hline
\hline
Parameter        &  Value           \\
\hline
$\pi$            & $19.99 \pm 0.38$  \\
$\alpha$     &  0.367$\pm$0.025 \\
M$_V$ (mag)           & $2.105 \pm 0.045$  \\
M$_{\rm bol}$  (mag)  & $2.14 \pm 0.04$  \\
T$_{\rm eff}$ (K)   & $7810 \pm 90 $   \\
$\log(L/L_\odot)$ (dex) & $1.04 \pm 0.02$   \\
$\log g$   (dex)      & $4.17 \pm 0.06$   \\
R/R$_\odot$      & $1.97 \pm 0.14$   \\
$M/M_\odot$      & $1.77^{+0.12}_{-0.19}$   \\
age (Myrs)       & $790 \pm 90  $   \\
\hline
\hline
\end{tabular}
\end{table}

Another very important piece of information is represented by the
angular diameter of  HD\,71297, which was published by
\citet{lafrasse10}: $\alpha$\,=\,0.367\,$\pm$\,0.025~mas. 
By using this value, together with the distance derived from
Hipparcos parallaxes, d\,=\,50\,$\pm$\,1~pc, it is straightforward to 
evaluate the radius of our target: R\,=\,1.97$\pm$0.14 R$_\odot$. 

With the observational data available for HD 71297, the most convenient way to estimate the
effective temperature is from the definition of surface brightness as given in 
Eq. 14.8 of \citet{gray05}. After simple algebra, we get:

\begin{equation}
 \log T_{\rm eff} = c - 0.1\cdot V + 0.1\cdot BC - 0.5 \log \theta_R
\label{teff}
\end{equation}

\noindent
where $\theta_R$ is the angular radius in arcsec and c is a constant given by:

\begin{equation}
c = \log T_{\rm eff}^\odot + 0.1\cdot m_V^\odot -0.1\cdot BC_\odot + 0.5 \log \theta_R^\odot
\end{equation}
 
To estimate $c$ we used the Sun ``constants'' reported by
\citet{bessel98}: $T_{\rm eff}^\odot$\,=\,5781$\pm$4 K, 
$m_V^\odot$\,=\,$-$26.76$\pm$0.01 mag, $BC_\odot$\,=\,$-$0.07$\pm$0.01 mag, 
and $\theta_R^\odot$\,=\,959.61$\pm$0.01 arcsec. Thus c\,=\,2.58405$\pm$0.00140,
and, from Eq.~\ref{teff}: $T_{\rm eff}$\,=\,7850\,$\pm$\,280 K. In the
error budget the most important contributor is the angular radius
(diameter) at the $\sim$3.5\% level, being the contribution of the
other quantities practically negligible. The effective temperature
value derived here is almost indistinguishable from that estimated
directly from Str\"omgren and Geneva photometry, but with a larger
uncertainty.  Nevertheless, this evaluation appears robust, since it is based on
different kind of  independent measures (photometry,
interferometry). Moreover, also the uncertainty can be calculated in a
reliable and physical way.  Hence, we decided to take a weighted
average of the effective temperatures obtained with the two methods as
the value to be used as a 
reference for comparison with spectroscopy. As a results, we obtain 
$T_{\rm eff}=7810\pm90$K. 

To estimate $\log g$, it is convenient to use the following expression
that can be easily derived from the definition of $g$ and from the
Stefan-Boltzmann relation:

\begin{eqnarray}
logg &= &4\log (T_{\rm eff} /T_{\rm eff}^\odot) + \log (M/M^\odot) +2 \log (\pi)
\nonumber \\
&&+0.4 (V+ BC_V + 0.26) + 4.44 
\label{logg}
\end{eqnarray}

Where the various terms of the equation have the usual meaning and
$M/M^\odot$ is the ratio mass of the star over mass of the Sun.
Before using Eq.~\ref{logg}, we have to evaluate the mass of
HD\,71297. This can be safely done by adopting the calibration 
mass--$M_V$ (valid for luminosity class V stars) by \citet{malkov03} 
that was derived on the basis of a large sample of eclipsing binaries stars. 
Hence, by using our $M_V$ estimate, we evaluated $\log
(M/M^\odot)=0.248\pm$ 0.05 dex (or $M/M^\odot=1.77^{+0.12}_{-0.19}$),  
where the error is largely conservative and completely dominated by the dispersion of the
mass--$M_V$ relation. 

Finally, applying Eq.~\ref{logg} we obtain $\log
g$\,=\,4.17\,$\pm$0.06, again,  in excellent agreement with the
purely photometric evaluation but with a much smaller uncertainty. We
will then use this value for comparison with spectroscopy.

\begin{figure}
\centering
\includegraphics[width=8.5cm]{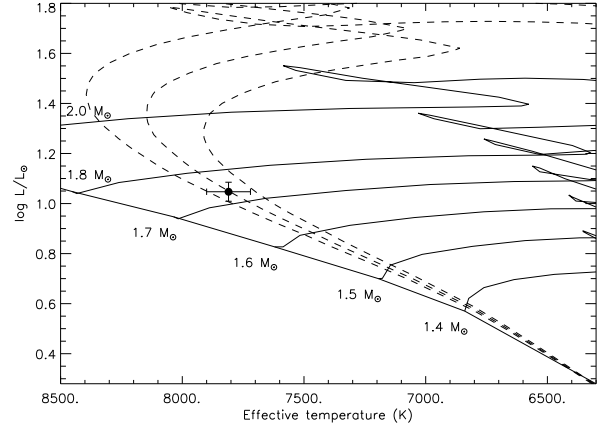}
\caption{Location of HD\,71297 in the HR diagram together with evolutionary
tracks and isochrones for $\log t$ ranging from 8.85 to 8.95 (step 0.05 and $t$ in yrs).}
\label{HR}
\end{figure}

Before closing this section, for completeness, it is worth estimating
the age of HD\,71297, taking advantage of the $\log T_{\rm eff}$ and $\log(L/L_\odot)$ 
values estimated in this section. 
The location of the star in the HR diagram, together with some evolutionary 
tracks computed for the solar metallicity Z\,=\,0.019 \citep{girardi00}, is shown in Fig.~\ref{HR}.  
Also shown are isochrones computed by \citet{marigo08} for the same Z (=\,0.019) and for ages ranging
from $\log t$\,=\,8.85 to 8.95, in steps of 0.05 ($t$ in years). The location of the star indicates a mass of 
M\,$\approx$1.75\,$\pm$\,0.05~M$_\odot$ (in perfect agreement with the value adopted above) and an age of 
$t \approx$\,790\,$\pm$\,90~Myrs. 

In Table~\ref{phot} we summarized all the astrophysical quantities for HD\,71297
evaluated independently from spectroscopy.

\section{Atmospheric parameters from spectroscopy}

In this section we will present separately the results from the analysis
of the high resolution spectra of HD\,71297 as obtained with the two 
different approaches quoted in the introduction.
We want to stress that the two analysis have been performed by using the same spectrum, prepared as
described in Section~\ref{obs}. In this way, we can be sure that any differences arising during the 
analysis will not be attributed to the quality of the data, but directly to the method itself.

\begin{figure} 
\includegraphics[width=8.8cm]{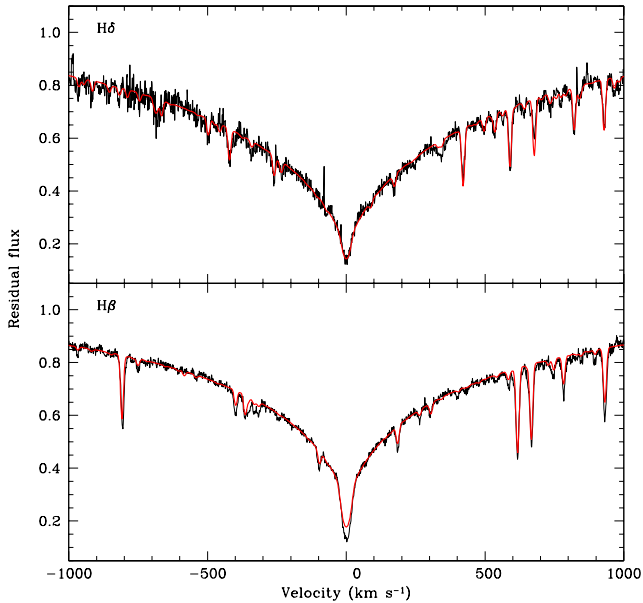} 
\caption{Observed H$_{\beta}$ and H$_{\delta}$ with over-imposed the synthetic ones computed with the fundamental 
parameters derived in this section.} 
\label{hbeta} 
\end{figure}

\subsection{Stellar parameters and abundance analysis with classical
  spectral synthesis}
\label{spec_synt}

This approach was already discussed in \citet{catanzaro11,catanzaro12}
and references therein. Thus, here we briefly summarize the principal
features of this method. 
In order to determine the optimal parameters, we minimize the 
difference between the observed and synthetic spectrum. Thus we minimize

\begin{displaymath}
 \chi^2 = \frac{1}{N} \sum \left(\frac{I_{\rm obs} - I_{\rm th}}{\delta I_{\rm obs}}\right)^2
\end{displaymath}

where $N$ is the total number of points, $I_{\rm obs}$ and $I_{\rm th}$ are the 
intensities of the observed and computed profiles, respectively, and $\delta I_{\rm obs}$ 
is the photon noise.  Synthetic spectra were generated in three  steps.   First, 
we computed a LTE atmospheric model using the ATLAS9 code \citep{kur93}. The stellar spectrum was 
then synthesized using SYNTHE \citep{kur81}.  Finally, the spectrum was convolved 
with the instrumental and rotational profiles.

As starting values of T$_{\rm eff}$ and $\log g$, we used the values derived
in the previous section.

To decrease the number of parameters, we computed the $v \sin i$ of 
HD\,71297 by matching synthetic line profiles from SYNTHE to a number of 
metallic lines.  The Mg{\sc i} triplet at $\lambda \lambda$5167-5183 {\AA} 
was particularly useful for this purpose.  The best fit was obtained for
$v \sin i$\,=\,7.0\,$\pm$\,0.5~km~s$^{-1}$.  This value is slightly lower than the one 
published by \citet{royer02} of $v \sin i$\,=\,11~km~s$^{-1}$.  

Uncertainties in T$_{\rm eff}$, $\log g$, and $v \sin i$ were estimated by the change in parameter values
which leads to an  increases of $\chi^2$ by unity \citep{lampton76}.

To determine stellar parameters as consistently as possible with the actual 
structure of the  atmosphere, we performed the abundance analysis by 
the following iterative procedure: 

\begin{description}

\item{(i)} $T_{\rm eff}$ is estimated by computing the ATLAS9 model atmosphere
which gives the best match between the observed H$_{\beta}$ and
H$_{\delta}$ lines profile and those 
computed with SYNTHE (see Fig.~\ref{hbeta}). The model was computed using solar opacity distribution 
function (ODF) and microturbulence 
velocity $\xi$\,=\,2.4\,$\pm$\,0.5~km~s$^{-1}$, the latter value has been calculated following the 
calibration $\xi\,=\,\xi(T_{\rm eff},\log g)$ published by \citet{allende04}.
These two lines are the only useful for our purpose since they are located
far from the echelle orders edges so that it was possible to safely recover the whole profiles.
The simultaneous fitting of two lines led to a final solution as the intersection of the 
two $\chi^2$ iso-surfaces. 
Another source of uncertainties is due to the difficulties in normalization as is always
challenging for Balmer lines in echelle spectra. We quantified the error introduced by the 
normalization in at least 100~K, that we summed in quadrature with the error obtained by
the fitting procedure. The final result is:

\begin{displaymath}
     T_{\rm eff}\,=\,7500\,\pm\,180\,K
\end{displaymath}

\noindent

\begin{figure} 
\includegraphics[width=8.8cm]{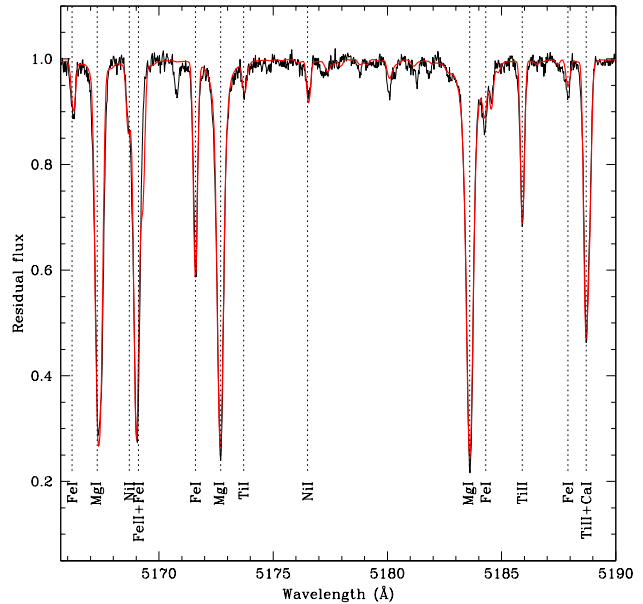} 
\caption{Observed Mg{\sc i} triplet with over-imposed 
the synthetic one computed with the fundamental parameters derived in this section} 
\label{mg} 
\end{figure} 

\begin{figure*}                                       
\includegraphics[width=15cm,angle=0]{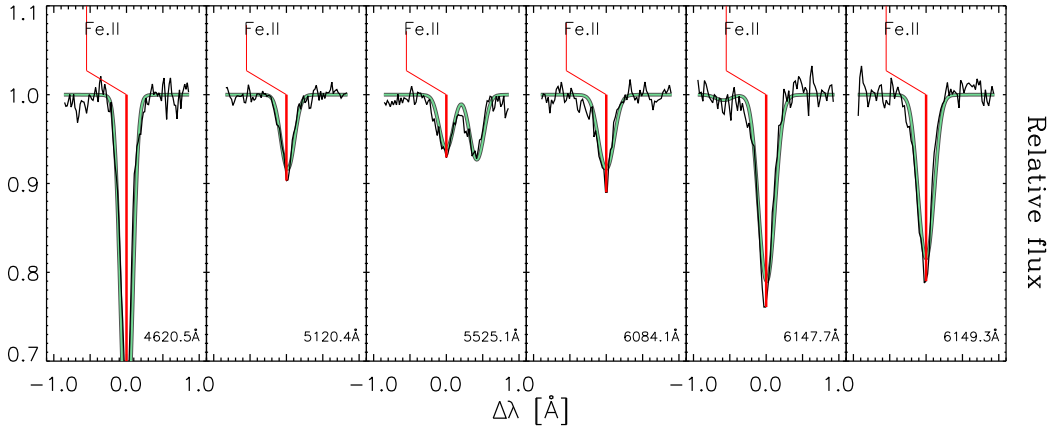} 
\caption{The figure shows six Fe{\sc ii} lines in HD~71297 fitted by {\sl VWA} (continuous line). The wavelengths 
of the fitted lines are given in the bottom right corner of each panel.} 
\label{lines} 
\end{figure*} 
The surface gravity was estimated from line profile fitting of Mg{\sc i} lines with developed wings.
This method is based on the fact that the wings of the Mg{\sc i} triplet at $\lambda\lambda$~5167, 5172, and 5183 {\AA}
lines are very sensitive to $\log g$ variations. In practice, we have first derived the magnesium abundance 
through the narrow Mg{\sc i} lines at $\lambda \lambda$~4571, 5528, 5711~{\AA} and the Mg{\sc ii} $\lambda$~7877~{\AA}, 
and then we fitted the wings of the triplet lines by fine tuning the $\log g$ value. The magnesium abundance we got from these
lines was $\log Mg/N_{\rm tot} = -4.62 \pm 0.14$.

To derive $\log g$ by fitting spectral wings is essential to take into 
account very accurate measurements of the atomic parameters of the transitions, i.e. $\log gf$ and the radiative,
Stark and Van der Waals damping constants. Regarding $\log gf$ we used the values of \citet{aldenius07}, Van der Waals
damping constant is that calculated by \citet{barklem00} ($\log \gamma_{\rm Waals}\,=\,-7.37$), the Stark damping constant 
is from \citet{fossati2011} ($\log \gamma_{\rm Stark}\,=\,-5.44$), and the radiative damping constant is from NIST database
($\log \gamma_{\rm rad}\,=\,7.99$). 

This procedure results in the final value of:

\begin{displaymath}
 \log g\,=\,4.0\,\pm\,0.1. 
\end{displaymath}

\item{(ii)} As a second step we determine the stellar abundances by computing a synthetic 
spectrum that reproduced the observed one. Therefore, we divide our spectrum into several 
intervals, 25~{\AA} wide each, and derived the abundances 
in each interval by performing a $\chi^2$ minimization of the difference between 
the observed and synthetic spectrum. The minimization algorithm has been written in {\it IDL}
language, using the {\it amoeba} routine. We adopted lists of spectral lines and atomic 
parameters from \citet{castelli04}, who updated the parameters listed originally by \citet{kur95}.

\end{description}

For each element, we calculated the uncertainty in the abundance to be the standard 
deviation of the mean obtained from individual determinations in each interval of the 
analyzed spectrum. For elements whose lines occurred in one or two intervals only, 
the error in the abundance was evaluated by varying the effective temperature and 
gravity within their uncertainties given in Table~\ref{abund}, 
$[ T_{\rm eff}\,\pm\, \delta T_{\rm eff}]$ and $[\log g\,\pm\,\delta \log g]$, and 
computing the abundance for $T_{\rm eff}$ and $\log g$ values in these ranges.
We found a variation of $\approx$~0.1 dex due to temperature variation. We did not
find any significant abundance change by varying $\log g$. The uncertainty in the temperature
is the main error source on our abundances. 

\begin{figure}                                       
\includegraphics[width=9cm]{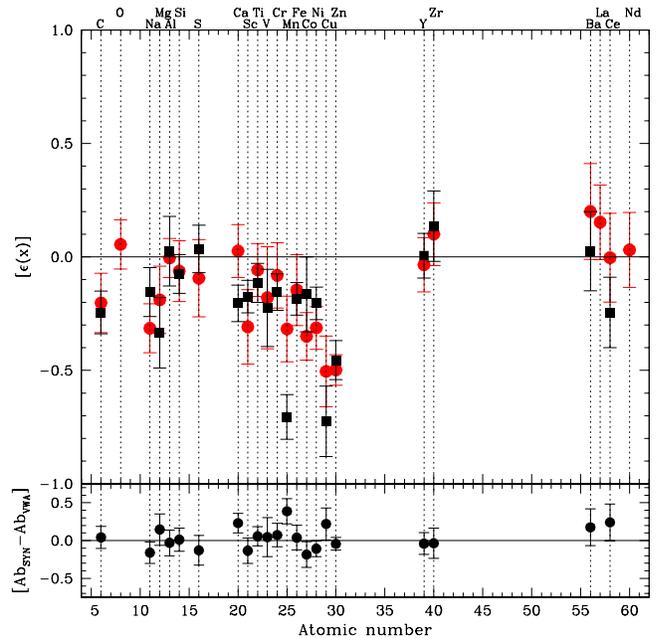} 
\caption{Chemical pattern for HD\,71297 derived with the two different methods, red dots are from spectral synthesis
and black squares from {\sl VWA}. The bottom panel shows the difference between the two methods, for all the chemical 
elements in common.} 
\label{pattern} 
\end{figure}

The derived abundances, expressed in terms of solar values \citep{grevesse10}, are shown in 
Fig.~\ref{pattern} (red dots). For all the chemical elements for which we detected spectral 
lines in our spectrum, we found moderate under-abundances with respect to the solar analogues, with 
the exception of the elements with high Z ($\ge$~40), for which we found normal abundances.

As an example of our spectral synthesis we show in Fig.~\ref{mg}, the comparison
between the observed spectral range from 5160 {\AA} to 5190 {\AA} and the synthetic one.
In the second column of Table~\ref{abund}, we listed the abundances for our star derived 
with the method described above.

\subsection{Stellar parameters and abundance analysis with {\sl VWA}}
\label{vwa}
The software package {\sl VWA} relies on the iterative fit of synthetic profile
regions around reasonably isolated absorption lines. {\sl VWA} has a graphical user interface (GUI),
which allows the user to investigate the spectra in detail, pick lines manually, 
inspect the quality of fitted lines, etc.

{\sl VWA} uses data from various sources. In particular, for the 
calculation of synthetic spectra it uses the $SYNTH$ code by
\citet{sme}, which works with $ATLAS9$ models, and atomic parameters 
and line-broadening coefficients from the $VALD$ database \citep{vald}.
We note that we used model atmospheres interpolated in the fine grid published by \citet{heiter02}. 
These models rely on the original ATLAS9 code by \cite{kur93} 
but use a more advanced convection description \citep{kupkaconv} based
on the works by \citet{canuto92}.

A thorough description of how {\sl VWA} works can be found in \citet{bruntt04,bruntt08}. Here we only recall 
the main steps of the analysis.

\begin{figure}                                       
\includegraphics[width=8.8cm,angle=0]{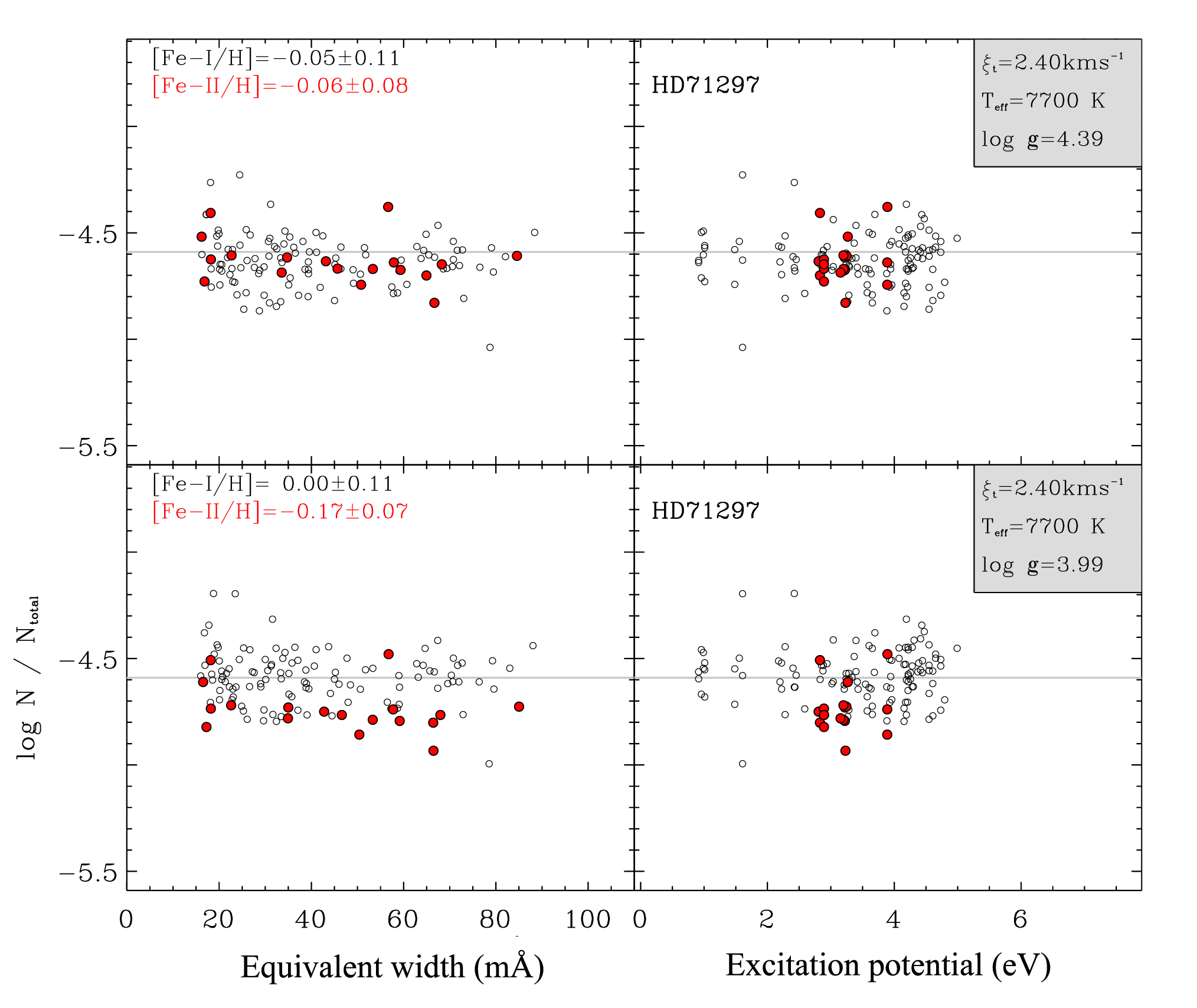} 
\caption{Abundance of Fe in HD~71297 for two different atmosphere
  models as a function of equivalent width (left panels) and
  excitation potential (right panels). Fe{\sc i} and Fe{\sc ii} lines
  are plotted with open and solid points, respectively. 
  The horizontal line in each
  panel show the  value of log N$_{\rm Fe}$/N$_{\rm total}$=-4.53
  adopted for the Sun, according to \citet{grevesse10}.
Top panels show the results for the adopted (best) parameters, while bottom panels
display the effect of decreasing $\log g$ by 0.4 dex (see labels).
} 
\label{vwares} 
\end{figure}

\begin{itemize}

\item {\it Normalization}: the first step consists in an accurate normalization of the spectrum. This can 
be achieved by adopting a synthetic spectrum (with approximately the same stellar parameters as the target) 
as a template to properly identify (in the GUI) suitable regions to anchor the continuum of the observed spectrum. 

\item {\it Setup and line selection}: atomic parameters and a preliminary
  model are setup on the basis of initial values for the stellar
  parameters. The line selection can be carried out either in 
automatic or manual way. We first searched for good lines (i.e. with
low blending)  in an automatic way, then we checked the result
visually line by line on the
GUI. 

\item {\it Fit of the lines and check of the result}: each line is fitted by iteratively 
changing the abundance to match the 
equivalent width (EW) of the observed and calculated spectrum.
The fitted lines are inspected in the GUI, problems with 
the continuum level or asymmetries in the line 
are readily identified, and these lines are discarded.  
This is done automatically by calculating the $\chi^2$ of the
fit in the core and the wings of the lines. This is followed by
a manual inspection of the fitted lines. Figure~\ref{lines} displays 
a few selected lines and the relative fits to demonstrate the quality
of the observed data (narrow lines) and the fit with {\sl VWA} (thick lines).

\item {\it Iterative estimate of T$_{\rm eff}$, $\log g$, and $\xi$}: 
The atmospheric parameters and the microturbulence
were then refined in several steps. This was done by minimizing the
correlations between the abundance of [Fe{\sc i}/H]  lines and equivalent
width and excitation potential (EP), and requiring good agreement between
the abundances of [Fe{\sc i}/H] and [Fe{\sc ii}/H] (the
difference in abundance: A(Fe{\sc i}) -A(Fe{\sc ii}) is often referred
as ``ionization balance''). This task is
accomplished by inspecting the results on the GUI and calculating the
needed models as required (see Fig.~\ref{vwares}).

\item {\it Chemical abundance}:
on the basis of the atmospheric parameters determined from [Fe{\sc i}/H] and
[Fe{\sc ii}/H], the mean abundances for all the elements with good lines 
can be finally calculated. 

\end{itemize}

In Fig.~\ref{vwares} we show the abundances of Fe for the best model
(T$_{\rm eff}$=7700\,$\pm$\,150 K; $\log g$=4.39\,$\pm$\,0.06 dex;
$v \sin i$= 7.0\,$\pm$\,1  km s$^{-1}$ and $\xi$=2.4\,$\pm$\,0.2 km s$^{-1}$)
and for a model with $\log g$ decreased by 0.4 dex, i.e. a value
approximately equal to that evaluated above with the Mg{\sc i}
triplet. The abundances are shown versus EW and EP. Open circles are
Fe{\sc i}  and solid circles are Fe{\sc ii} lines, respectively. The mean abundance and rms scatter
of Fe{\sc i}  and Fe{\sc ii} lines are given in each panel. We note that in the calculation of the
ionization balance we adopt the rms of the mean, hence the uncertainties on
the Fe{\sc i}  and Fe{\sc ii} abundances are significantly smaller (e.g. $\sim$ 0.01-0.02 dex).
Bearing this in mind, a comparison of top and bottom panels in the figure clearly show that
in the {\sl VWA} framework a value $\log g \approx$ 4.0 dex is unlikely.

To estimate the uncertainties of the atmospheric parameters,
we repeated the above outlined analysis by varying significantly $T_{\rm eff}$, $\log g$ and
$\xi$ (one parameter is allowed to vary while the other two are kept fixed). 
In this way we can determine when the correlations with EW and EP become
significant or the ionization balance begins to
deviate from equality \citep[see][for details]{bruntt08}. 
The same perturbed models can be used to estimate the uncertainty on
the [Fe/H] value by estimating the variation on its value caused by 1\,$\sigma$
perturbation of one parameter among T$_{\rm eff}$, $\log g$, $\xi$ and
taking fixed the other two \citep[see][for a detailed
discussion]{bruntt08}. The resulting three uncertainties were summed
in quadrature together with the rms scatter of Fe{\sc i} to obtain
the final error on [Fe/H].
We underline that this method for the determination of the
uncertainties gives only an internal estimate since
the absolute temperature scale of the model atmospheres may be
systematically wrong. Thus, our measures could show a
good precision, but the accuracy is most likely not as good \citep{bruntt10b}.

It is important to note that one of the physical assumptions in the
models adopted here is local thermodynamical equilibrium (LTE), 
but deviations from LTE start to become important for stars hotter than
about 6300 K, especially for stars more metal poor than the Sun.
Thus we have included the NTLE corrections in
the present analysis according to \cite{rent96}. 
The correction for neutral iron in our case is 
[Fe{\sc i}/H]$_{\rm NLTE}=$\,[Fe{\sc i}/H]$_{\rm LTE}+0.11$ dex. 
When this correction is applied, Fe{\sc ii} (unaffected by NLTE) must be 
increased by adding $+0.2$ to $\log g$. This correction has been applied in the
results from {\sl VWA} reported here.

The stellar parameters and the abundances for iron and the other
measured chemical species obtained by means of {\sl VWA} are shown in
Table~\ref{abund}. The uncertainties were calculated as for iron but
only for elements with at least three lines measured. For those chemical species for which less than three lines
have been identified, we arbitrarily set the errors in 0.15~dex for elements with only one line and
0.10 dex for elements with two lines.

\begin{table}
 \centering
  \caption{Comparison among atmospheric parameters and abundances derived by spectral synthesis modeling and
           by {\sl VWA} approach. $N$ denotes the number of lines used with
           {\sl VWA} package. T$_{\rm eff} $ is in Kelvin, $\log g$ is in
           dex, while  $v \sin i$ and $\xi$ are in Km/s. Abundances
           are expressed in the form $\log N_{\rm el}/N_{\rm Tot}$. An
         asterisk indicate that these uncertainties were re-determined
         in Sect. 6.}
  \begin{tabular}{lccrc}
  \hline
  \hline
      & Synthesis & {\sl VWA} & N & Sun \\
 \hline
 T$_{\rm eff} $         & 7500\,$\pm$\,180  &  7700\,$\pm$\,150   &  & \\
 $\log g$                 & 4.0\,$\pm$\,0.28$^{*}$   &  4.39\,$\pm$\,0.20$^{*}$  &  & \\
 $v \sin i$ & 7.0\,$\pm$\,0.5   &  7.0\,$\pm$\,1.0    &  & \\
 $\xi$     & 2.4\,$\pm$\,0.5           &  2.4\,$\pm$\,0.2    &  & \\ 
\hline
  C   &  $-$3.81\,$\pm$\,0.12 &$-$3.85\,$\pm$\,0.08  & 7  & $-$3.60\,$\pm$\,0.05 \\
  O   &  $-$3.29\,$\pm$\,0.10 &  ---                 & -  & $-$3.34\,$\pm$\,0.05 \\
  Na  &  $-$6.11\,$\pm$\,0.10 &$-$5.95\,$\pm$\,0.10  & 2  & $-$5.79\,$\pm$\,0.04 \\
  Mg  &  $-$4.62\,$\pm$\,0.14 &$-$4.77\,$\pm$\,0.15  & 1  & $-$4.43\,$\pm$\,0.04  \\
  Al  &  $-$5.59\,$\pm$\,0.08 &$-$5.56\,$\pm$\,0.15  & 1  & $-$5.58\,$\pm$\,0.03  \\
  Si  &  $-$4.59\,$\pm$\,0.13 &$-$4.60\,$\pm$\,0.08  & 8  & $-$4.52\,$\pm$\,0.03    \\
  S   &  $-$5.00\,$\pm$\,0.17 &$-$4.88\,$\pm$\,0.10  & 2  & $-$4.91\,$\pm$\,0.03    \\
  Ca  &  $-$5.62\,$\pm$\,0.15 &$-$5.90\,$\pm$\,0.07  & 9  & $-$5.69\,$\pm$\,0.04    \\
  Sc  &  $-$9.19\,$\pm$\,0.16 &$-$9.06\,$\pm$\,0.06  & 4  & $-$8.88\,$\pm$\,0.04    \\
  Ti  &  $-$7.14\,$\pm$\,0.10 &$-$7.20\,$\pm$\,0.07  & 24 & $-$7.08\,$\pm$\,0.05    \\
  V   &  $-$8.28\,$\pm$\,0.21 &$-$8.33\,$\pm$\,0.15  & 1  & $-$8.10\,$\pm$\,0.08    \\
  Cr  &  $-$6.48\,$\pm$\,0.14 &$-$6.55\,$\pm$\,0.07  & 26 & $-$6.39\,$\pm$\,0.04    \\
  Mn  &  $-$6.92\,$\pm$\,0.14 &$-$7.31\,$\pm$\,0.09  & 7  & $-$6.60\,$\pm$\,0.04    \\
  Fe  &  $-$4.68\,$\pm$\,0.15 &$-$4.72\,$\pm$\,0.06  & 125 & $-$4.53\,$\pm$\,0.04    \\
  Co  &  $-$7.39\,$\pm$\,0.08 &$-$7.21\,$\pm$\,0.15  & 1  & $-$7.04\,$\pm$\,0.07    \\
  Ni  &  $-$6.13\,$\pm$\,0.08 &$-$6.02\,$\pm$\,0.06  & 26 & $-$5.81\,$\pm$\,0.04    \\
  Cu  &  $-$8.35\,$\pm$\,0.15 &$-$8.57\,$\pm$\,0.15  & 1  & $-$7.84\,$\pm$\,0.04    \\
  Zn  &  $-$7.97\,$\pm$\,0.04 &$-$7.93\,$\pm$\,0.07  & 2  & $-$7.47\,$\pm$\,0.05    \\
  Y   &  $-$9.86\,$\pm$\,0.11 &$-$9.82\,$\pm$\,0.09  & 4  & $-$9.82\,$\pm$\,0.05    \\
  Zr  &  $-$9.35\,$\pm$\,0.13 &$-$9.32\,$\pm$\,0.15  & 1  & $-$9.45\,$\pm$\,0.04    \\
  Ba  &  $-$9.65\,$\pm$\,0.19 &$-$9.83\,$\pm$\,0.15  & 1  & $-$9.85\,$\pm$\,0.09    \\
  La  & $-$10.78\,$\pm$\,0.16 &  ---                 & -  & $-$10.93\,$\pm$\,0.04    \\
  Ce  & $-$10.46\,$\pm$\,0.19 &$-$10.70\,$\pm$\,0.15 & 1  & $-$10.45\,$\pm$\,0.04    \\
  Nd  & $-$10.58\,$\pm$\,0.16 &  ---                 & -  & $-$10.61\,$\pm$\,0.04    \\
 \hline
\end{tabular}
\label{abund}
\end{table}

\section{Comparison of the methods}
\label{disc}
In the previous section we have analyzed the high resolution spectrum of the Am: star
HD\,71297 using two different approaches: the classical spectral
synthesis and {\sl VWA}. The results
for effective temperature, gravity, $v \sin i$, microturbulence, and
chemical abundances, obtained with the two methods have been
summarized in Tab.~\ref{abund}. We can now compare between them the stellar
parameters obtained with the two quoted methods, taking also into
account the completely independent results obtained in
Sect.~\ref{parallax} for T$_{\rm eff}$ and $\log g$ as summarized in Tab.~\ref{phot}. 

\begin{figure}                                       
\includegraphics[width=9cm]{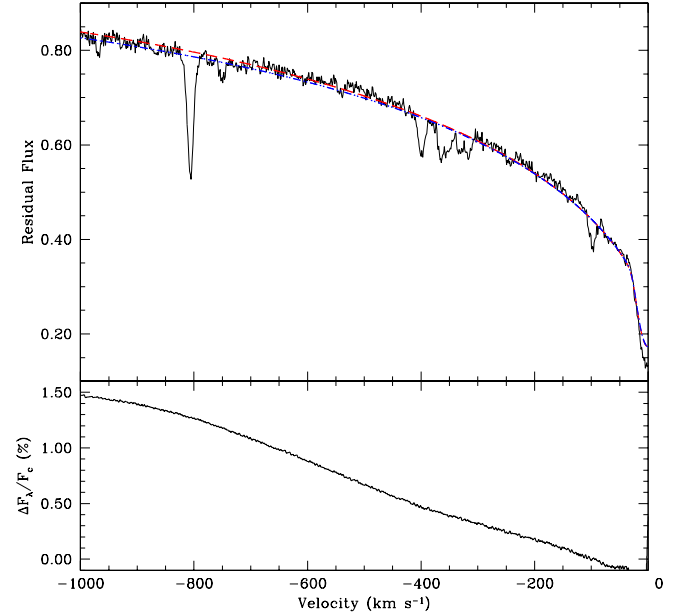} 
\caption{{\it Upper panel} - Comparison between observed and synthetic H$\beta$ profiles computed
with two convection models represented by red dashed line (MLT) and by blue dot-dashed line (CM).
{\it Bottom panel} shows the differences between normalized fluxes in H$\beta$ profile obtained from
two models differing only by the convection treatment.} 
\label{comp_hb} 
\end{figure}

First of all, the atmospheric models we used in our calculations are both ATLAS9
based but with different convection zone treatments. In particular, in Sect.~\ref{spec_synt}
it was computed using the classical treatment MLT with fixed $\alpha$=1.25 \citep{castelli97}. 
In Sect.~\ref{vwa}, the model was interpolated in the fine grid by \citet{heiter02}, which uses 
a more advanced convection description based on \citet{canuto92} (CM).  

\citet{heiter02} investigated the effects of the different convection approaches into the 
Balmer lines profiles and they concluded that the observed profiles should have an accuracy of at
least 0.5\,$\%$ to clearly appreciate the differences between convection models with different
efficiency. 

In order to quantify these effects on our target, we computed two synthetic H$\beta$
profiles using the same T$_{\rm eff}$\,=\,7700~K and $\log g$\,=\,4.39 but with different 
convection models. We show this comparison in Fig.~\ref{comp_hb}, where in the upper panel we compare
the observed blue wing of the H$\beta$ with overimposed the profiles computed with MLT (red dashed 
curve) and with CM treatment (blue dot-dashed line). In the bottom panel
we report the difference in percentage between the two models. Since the maximum difference is of 
the order of 1.5\,$\%$, indistinguishable at our level of signal-to-noise ratio, we conclude that 
for our target we can neglect the differences arising from the convection models adopted. 

To verify the accuracy of this conclusion we have repeated the calculation for effective temperature, gravity,
and abundances as described in Sect.~\ref{spec_synt}. Again, we estimated T$_{\rm eff}$ by simultaneous
fitting of H$\gamma$ and H$\delta$ but using ATLAS9 models modified for the CM treatment convection.
We obtained T$_{\rm eff}$\,=\,7600\,$\pm$\,150~K, that is closer to
the photometric/interferometric value, but also totally 
consistent with the value derived adopting the MLT treatment for
convection. Using this temperature, we estimated gravity fitting the Mg{\sc i}
triplet and we obtain $\log g$\,=\,4.10\,$\pm$\,0.10, again compatible with previous value.

\begin{figure}                                       
\includegraphics[width=9cm]{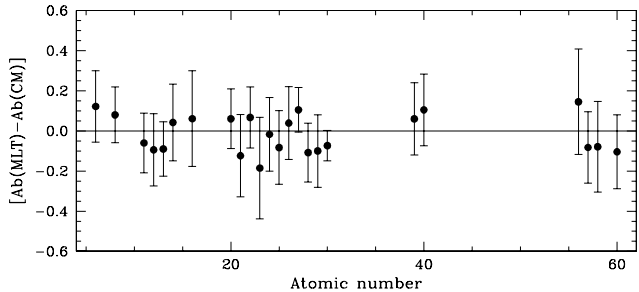} 
\caption{Comparison between abundances computed using models that differ for the treatment of the convection
theory, MLT vs CM.} 
\label{MLTvsCM} 
\end{figure}

For what concerns the abundances, the model built with these values of effective temperature and gravity has been
used to repeat the spectral synthesis analysis, adopting the same intervals as in Sect.~\ref{spec_synt}. 
In Fig.~\ref{MLTvsCM} we displayed the differences between the
abundances reported in Tab.~\ref{abund} computed with MLT based model and those computed with CM model.
Also in this case the results are consistent, being the weighted average of all differences
equal to $-$0.01\,$\pm$\,0.09. 

Starting with T$_{\rm eff}$, the classical spectral synthesis method and {\sl VWA}
agree well within 1\,$\sigma$. There is also a 1.3\,$\sigma$ agreement
with the T$_{\rm eff}$  value estimated from photometry and interferometry, albeit
the {\sl VWA} value is closer by 200 K to the reference value. On the
contrary, concerning the value of $\log g$, there is a significant
discrepancy between the classical spectral synthesis method and {\sl VWA} results:  
$\log g$\,=4.0\,$\pm$\,0.1 dex from synthesis of the strong Mg{\sc i}
triplet and $\log g$\,=4.39\,$\pm$\,0.06 dex from ionization equilibrium 
of Fe{\sc i}/Fe{\sc ii} ({\sl VWA}). The situation is even more complex if
we take as a reference the gravity evaluated independently from
spectroscopy in Sect.~\ref{phot}: $\log g$=4.17$\pm$0.05 dex. 
This value lies halfway between the two spectroscopy-based
methods, being incompatible by more than 1 and 2\,$\sigma$ with respect
to the results of classical spectral synthesis and {\sl VWA} methods, respectively. 
Thus, if we trust in interferometry and parallax we have to conclude
that even high resolution spectroscopy is unable, at least in the case
HD\,71297, of evaluating the $\log g$ value better than 0.2 dex, being
also the internal errors on this value underestimated for both
methods. 

\begin{figure*} 
\includegraphics[width=18cm,bb=54 460 592 720]{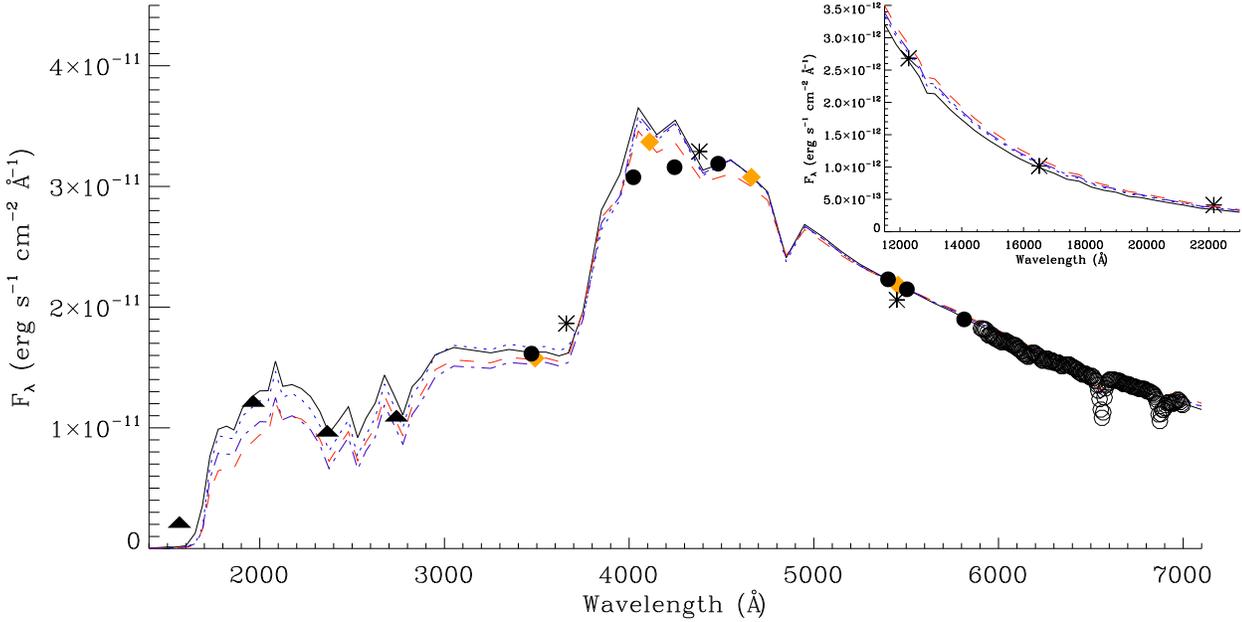} 
\caption{Comparison between observed spectral energy distribution and four theoretical fluxes. The meaning 
of the symbols is the following: triangles are TD1 fluxes, filled circles represent Geneva photometry, diamonds 
are fluxes from uvby magnitudes, asterisks are UBVJHK fluxes, and open circles represent the spectrophotometry.  
The theoretical distributions have been calculated for: T$_{\rm eff}$=\,7810 K and $\log g$=\,4.17 (continuous black line),
T$_{\rm eff}$=\,7700 K and $\log g$=\,4.39 (CM model, dotted blue line), T$_{\rm eff}$=\,7500 K and $\log g$=\,4.00 
(MLT model, dashed red line), and T$_{\rm eff}$=\,7600 K and $\log g$=\,4.10 (CM model, dash-dotted purple line)} 
\label{sed} 
\end{figure*} 

To investigate further the accuracy of the results obtained with the
two different spectroscopic analysis about  T$_{\rm eff}$ and $\log
g$, we built the Spectral Energy Distribution (SED) of HD\,71297 
(see Appendix for all the details on the construction of the SED) as shown in Fig.~\ref{sed}. The figure 
shows the comparison of the various photometric or spectrophotometric sources with three models (in
different colors and line styles), as estimated in the previous sections\footnote{All the models 
were calculated for [Fe/H]=$-$0.15 as derived for HD\,71297}. 
As well known, the Balmer jump is sensitive to surface gravity (as well as to effective temperature 
and line blanketing), hence looking at the UV part of the SED we can get some insight about the ``best'' couple 
(T$_{\rm eff}$, $\log g$). Even a qualitative check on the figure show that the T$_{\rm eff}$, $\log g$ couple  
evaluated in Sect.~\ref{phot_sect} (from parallax and interferometry) reproduces better the SED at all wavelength. 
In particular, in the far UV, the classical spectral synthesis method has the worse agreement due to the low T$_{\rm eff}$, 
whereas across the Balmer Jump, {\sl VWA} seems to have a too high $\log g$. This reflects the fact that both 
spectral synthesis and {\sl VWA} methods go through a modeling that introduces errors, then it is important to discuss
some of them.
  
The error we found in $\log g$ by fitting the wings of the Mg{\sc i} triplet (i.e. 0.1~dex)
is the formal error due to the fitting procedure. Actually there is at least another source of error
that we have to discuss and consider. The result of the modeling of a spectral line depends essentially
on the atomic parameters of the transition and in particular, for what concern the width of the spectral lines,
we have to pay attention to the radiative, Stark and Van der Waals damping constants. The values
of these three numbers set the broadening of the line and it is straightforward to understand 
the importance of their accuracy on the final value of $\log g$.

To focus our attention on this problem, we performed the following simulation: we fitted the observed
profile of the Mg{\sc i} $\lambda$5172.684 {\AA} on synthetic profiles by varying the damping constants,
one at a time, by an amount equal to their typical uncertainties. 
According to \citet{barklem00}, we adopted 5$\%$ as typical error in
damping constants, obtaining that such a variation of $\gamma_{\rm Stark}$ and $\gamma_{\rm Waals}$ lead to an 
uncertainty in $\log g$ equal to $\approx \pm$\,0.3\,dex. Similarly, a variation of $\gamma_{\rm rad}$ of 5\%, 
lead to a $\approx \pm$\,0.15\,dex uncertainty on $\log g$. Therefore, considering these errors as random and independent, 
we can conclude that the final uncertainties on gravity estimated by fitting the wings of one line is given from their 
quadratic sum and then equal to 0.45~dex. However, since we have applied this method to all the Mg{\sc i} triplet,
the final uncertainty coming from atomic parameters is 0.26~dex. Finally, considering also the formal error on the fitting
procedure, we obtained the total error on $\log g$ equal to 0.28~dex. 
With such an error, $\log g$ derived with classical spectral synthesis is consistent with other values 
found in this study.

 As for {\sl VWA}, the estimate of $\log g$ by means of the ionization balance can be affected by over-ionization 
effects and uncertainties in the temperature structure of the model atmosphere, as well as to
uncertainties on the atomic parameters \citep{fuhrmann97}. Thus, the
formal error on $\log g$ we derived in our analysis has to be regarded
as a lower limit for the true uncertainty on this parameter. 
Since it is difficult to estimate quantitatively the effect of the various
source of uncertainty on the final value of $\log g$, we decided to
assign a ``typical error'' of 0.2 dex. 
This suggestion is justified by the work of \citet{bruntt2012} who 
used {\sl VWA} to analyze high resolution spectra for 93 solar-type
stars possessing accurate $\log g$ values measured by means of 
asteroseismological techniques on the basis of outstanding Kepler
satellite photometry. These authors find an average difference
$\log g({\rm spec})-\log g({\rm asteroseismic})$=+0.08$\pm$0.07 dex 
with a few objects showing larger discrepancies ($\sim$0.3 dex). Thus, 
an uncertainty of the order of 0.2 dex in $\log g$ is absolutely
normal. A similar conclusion was reached by \citet{Smalley05}: ``Realistically, the
typical errors on the atmospheric parameters of
a star, will be T$_{\rm eff}\pm$100 K (1$\sim$2\%) for 
$\pm$0.2 dex ($\sim$20\%) for $\log g$ ".
Finally, as for the values of $v \sin i$, $\xi$, and chemical abundances, an inspection of Tab.~\ref{abund} reveals that 
there is a good agreement between spectral synthesis and {\sl VWA}, in
spite of the discrepancies in T$_{\rm eff}$ and $\log g$. The
agreement in chemical abundance is not surprising. As argued by 
\citet{Smalley05}, the atmospheric parameters
obtained from spectroscopic methods
alone may not be consistent with the ``true'' values
as obtained by model-independent methods, but 
this is not necessarily important for abundance
analyses of stars.

\section{Conclusion}

In this paper we have tested two different methods commonly used in the recent literature to characterize the 
stellar atmospheres and their chemical pattern: classical spectral synthesis and {\sl VWA} package. This test was performed  
analyzing the spectrum of HD\,71297 carried out with SARG@TNG. This object has been classified by \citet{cowley68} as a 
suspected marginal metallic star, and later on \citet{abt85} assigned the spectral type of kA8hA9mF0.

From direct measurements of distance and diameter, we obtained for our target the following astrophysical parameters:
T$_{\rm eff}$\,=\,7810\,$\pm$\,90\,K and $\log g$\,=\,4.17\,$\pm$\,0.05. As a by-product we were also able to derive estimation 
of R/R$_\odot$\,=\,1.97\,$\pm$\,0.14, $M/M_\odot$\,=\,1.77$^{+0.12}_{-0.19}$, and age\,=\,790\,$\pm$\,90\,Myrs.  

For what concern the main part of our paper, i.e. the comparison of the two different approaches, we can summarize as
follows:

\begin{itemize}
 \item Classical spectral synthesis method gives us the following values: T$_{\rm eff}$ = 7500 $\pm$ 180 K and 
       $\log g$ = 4.10 $\pm$0.28 that are consistent with the previous values, at least within the errors. Projected 
       rotational velocity has been evaluated in 7.0\,$\pm$\,0.5~km~s$^{-1}$ and $\xi$\,=\,2.4\,$\pm$\,0.5~km~s$^{-1}$. 
       With these parameters, abundance analysis gives us a general underabundance of iron peak elements with the exception 
       of calcium that is solar in content. Solar abundances are also shown by oxygen, yttrium, zirconium, barium and rare earths.
       We also tested the influence of the convection in the calculation of the synthetic spectrum. 
       As in the case of VWA, we used \citet{heiter02} models to repeat the analysis, obtaining results totally consistent 
       with those obtained with MLT theory as treated in \citet{kur93} with no overshooting and $\alpha$\,=\,1.25. 
       Thus we conclude that, at least in the case of a star as hot as HD 71297, and for the resolution and SNR of our
       spectrum, the results are only slightly affected by the choice about the particular treatment of convection adopted 
       for the analysis. The same conclusion is shared by \citet{gardiner99}, who conclude that the MLT and CM models all give
       similarly reasonable results. On the contrary, for cooler targets the role of convection will be more significant and we 
       will take it into account properly.

 \item By using {\sl VWA} we obtained the following parameters: T$_{\rm eff}$ = 7700 $\pm$ 150 K and $\log g$ = 4.39 $\pm$0.06, a 
       projected rotational velocity of 7.0\,$\pm$\,1.0~km~s$^{-1}$ and $\xi$\,=\,2.4\,$\pm$\,0.2~km~s$^{-1}$. Considering the 
       experimental errors all these quantities are comparable with the ones derived with the others method. Also the abundances
       show a pattern similar to the one computed with classical spectral synthesis.
\end{itemize}

As a general conclusion we can state that the methods considered in this study to derive fundamental parameters useful
to characterize stellar atmosphere give consistent results, if we consider all the sources of experimental errors. 

An important result of this study concern the chemical pattern found for HD\,71297. Contrary to what is expected from the 
previous classification, our abundances (reported in Tab.~\ref{abund}) do not look like those of normal Am star
\citep[see Fig.\,5 in][for example]{catanzaro11}. In fact, iron-peak 
elements show moderate underabundances, as well as carbon and sodium. Heavy elements like yttrium, zirconium, barium and rare
earths, that usually in A-type stars display abundances greater than the Sun analogues \citep{erspamer03}, are quite normal in our
target.

The results shown in this paper, concernin the consistency of classical spectral synthesis and {\sl VWA} approaches 
to the analysis of stellar spectra, will allow us to confidently apply these codes to the other Am stars observed at SARG@TNG.

\section*{Acknowledgments} 
We thank the  referee, M. S. Bessell, for providing constructive
comments that  helped us in improving this paper. 
This research has made use of the SIMBAD database and VizieR catalogue access tool,
operated at CDS, Strasbourg, France.

\appendix

\section{Spectral Energy Distribution}

The observed SED has been obtained by merging various data sources collected from the literature. 
In particular, the stellar flux shown in Fig.~\ref{sed} was constructed by using the following data:

\begin{itemize}
\item UV fluxes taken from TD1 satellite \citep{thompson78} that cover the 1565-2740 {\AA} range;

\item uvby magnitudes from \citet{hauck98} and converted in physical units by using the calibrations 
given by \citet{gray98}; {\it y} magnitude has been derived from UBV color via the calibration of
\citet{mcswain05}

\item Geneva photometry taken from \citet{rufener88} and converted in fluxes by means of the calibrations
given by \citet{rufnic88};

\item UBV magnitudes taken from \citet{mermil91} and converted in fluxes using \citet{bessel98} calibrations;

\item spectrophotometry in the range 5800-8000 {\AA} taken from \citet{weaver95};

\item JHK magnitudes from 2MASS survey \citep{2mass}, converted in physical units by using \citet{bliek96}.

\end{itemize}

As discussed in Sect.~\ref{phot_sect}, we neglected the effects of interstellar reddening. In order to be consistent 
with the iron abundance we found in our previous analysis, each theoretical SED has to be calculated with 
an opacity ODF corresponding to a metalicity of [Fe/H] $= -0.15$. To achieve this goal, we computed for each couple
of (T$_{\rm eff}$, $\log g$) two synthetic fluxes, one for ODF=[0.0] and one for ODF=[-0.5] and then we interpolated
between them. 

\label{lastpage} 
 
\end{document}